\documentclass[11pt,a4paper,english,nofootinbib,,superscriptaddress]{revtex4}
\usepackage{lmodern}

\usepackage[T1]{fontenc}
\usepackage[latin9]{inputenc}
\usepackage{babel}
\usepackage{color}
\usepackage{cancel}
\usepackage{amsmath}
\usepackage{graphicx}
\usepackage{amssymb}
\usepackage{esint}
\usepackage[unicode=true, pdfusetitle,
bookmarks=true,bookmarksnumbered=false,bookmarksopen=false,
breaklinks=false,pdfborder={0 0 1},backref=false,colorlinks=false]
{hyperref}
\usepackage{amsmath}
\usepackage{amssymb}
\usepackage{float}

\def\barra#1{\not \!#1}
\def\b{\begin{equation}} \def\e{\end{equation}}
\def\bd{\begin{displaystyle}} \def\ed{\end{displaystyle}}
\def\ba{\begin{array}} \def\ea{\end{array}}

\def\bee{\begin{enumerate}}
	\def\eee{\end{enumerate}}

\def\ud{\mathrm{d}}

\def\dg{^{\dag}}

\def\1{\mbox{I\hspace{-.15em}1}}

\def\R{{\rm I\hspace{-.15em}R}}

\def\b{\begin{equation}}
\def\e{\end{equation}}
\def\bee{\begin{enumerate}}
	\def\eee{\end{enumerate}}

\makeatletter
\usepackage{latexsym}\usepackage{bm}

\makeatother

\begin{document}
	\title{De Sitter scalar-spinor interaction in Minkowski limit}
	
	\author{Y. Ahmadi}
	\email{ahmadi.pave@gmail.com} \affiliation{Department of Physics, Razi University, Kermanshah, Iran}
\begin{abstract}

\noindent \hspace{0.35cm}
The scalar-spinor interaction Lagrangian is presented by the Yukawa potential. In dS ambient space formalism, the interaction Lagrangian of scalar-spinor fields was obtained from a new transformation which is very similar to the guage theory. The interaction of massless minimally coupled scalar and spinor fields was investigated. The Minkowski limit of the massless minimally coupled scalar field and massive spinor field interaction in the ambient space formalism of de Sitter space time is calculated. The interaction Lagrangian and massless minimally coupled scalar field in the null curvature limit become zero and the local transformation in the null curvature limit become a constant phase transformation and the interaction in this limit become zero. The covariant derivative reduces to ordinary derivative too. Then we conclude that this interaction is due to the curvature of space time and then the massless minimally coupled scalar field may be a part of a gravitational field.
 
\end{abstract}

\maketitle


\section{Introduction}

Our universe is expanding by a positive acceleration and experimental data such as highly redshift observation of supernova Ia \cite{Riess, Perl}, galaxy clusters \cite{Henry 1,Henry 2} and cosmic microwave background radiation \cite{Nature}, confirm it. Therefore, our universe maybe described by the de Sitter (dS) metric in the large scale. Furthermore, the recently observational data from \cite{BICEP2} confirm that the early universe in a good approximation is dS universe, the inflationary epoch. Thus the construction of quantum field theory (QFT) in dS space-time is essential. In dS ambient space formalism, the efforts have been made to construct QFT in last years \cite{ta97,77,ta96,taazba,taro12,berotata,derotata}. Fortunately, because of the linearity of the action of dS group on ambient space formalism, the construction of QFT is very similar to the Minkowski space-time. An another advantage of ambient space formalism is the analyticity properties of two-point function (which proved by Bros et al. \cite{brgamo,brmo,brmo03}). These properties are fundamental basis for the interaction calculations. The vector-spinor fields interaction in dS ambient space formalism was investigated in previous paper and the $\cal S$ matrix of Compton scattering was obtained in this formalism \cite{jaahta2017}, then it was showed that the null curvature limit of this $\cal S$ matrix is match with Minkowski counterpart. 
 
In the standard model of QFT, the interaction Lagrangian usually  is constructed from gauge theory; except scalar-spinor fields interaction, which con not be obtained from group theoretical point of view. The scalar-spinor interaction Lagrangian is presented by the Yukawa potential. In dS ambient space formalism, the interaction Lagrangian of scalar-spinor fields was obtained from a new transformation \cite{higgs}. This approach is very similar to the guage theory. Also in \cite{ahjata2019} the interaction of massless minimally coupled ({\it mmc}) scalar and spinor fields was investigated. Here, we tried to obtaine the Minkowski limit of this interaction. The null curvature limit of {\it mmc} scalar-spinor interaction Lagrangian and {\it mmc} scalar two-point function discussed in this paper.

The organization of this article is as follow. We recall the spinor and {\it mmc} scalar interaction fields in dS ambient space-time in section \ref{interaction}. In section \ref{flat limit}, the scaler-spinor interaction is investigated in null curvature limit. Finally, the conclusion has been presented in section \ref{Conclusion}.

\setcounter{equation}{0}	
\section{the Interaction in $dS$ ambient space formalism}\label{interaction}

The dS space-time can be considered as a 4-dimensional  hyperboloid embedded in 5-dimensional Minkowski space with the following relation:
$$
M_H=\{x \in \R^5| \; \; x \cdot x=\eta_{\alpha\beta} x^\alpha
x^\beta =-H^{-2}\};\;\; \alpha,\beta=0,1,2,3,4 ,$$
where
 $\eta_{\alpha\beta}=\mbox{diag}(1,-1,-1,-1,-1)$, $H$ is Hubble constant parameter. The dS metric is:
$$
ds^2=\eta_{\alpha\beta}dx^{\alpha}dx^{\beta}|_{x^2=-H^{-2}}=g_{\mu\nu}^{dS}dX^{\mu}dX^{\nu};\;\; \mu,\nu=0,1,2,3 ,$$
that $X^\mu$ is dS intrinsic coordinate and $x^\alpha$ is the 5-dimensional ambient space formalism. There are many coordinate systems which represent the ambient space coordinate $x^\alpha$ in terms of the intrinsic coordinate $X^\mu$. For investigating the flat space limit, the following coordinate system is suitable:
  \b \label{flat cordinates}
x^\alpha= \left(
H^{-1}\sinh(HX^0),\;
H^{-1}\dfrac{\overrightarrow{X}}{\lVert\overrightarrow{X}\lVert}\cosh(HX^0)\sin(H\lVert\overrightarrow{X}\lVert),\;
H^{-1}\cosh(HX^0)\cos(H\lVert\overrightarrow{X}\lVert)\right)\e
where $\lVert\overrightarrow{X}\lVert=(X_1^2+X_2^2+X_3^3)^\frac{1}{2}$ is the norm of three-vector $\overrightarrow{X}$ \cite{brgamo}.
The Fourier transformation cannot be defined in curved space-time, but in the  dS space-time, the Fourier-Helgason-type transformation can be defined because of maximally symmetric properties of dS hyperboloid \cite{hel62,hel94,brmo03,brmo}.  In corresponding with any space-time variable $x^\alpha$, one can defined a $\xi^\alpha=(\xi^0, \vec \xi, \xi^4)$ in the positive null-cone $C^+=\left\lbrace \xi \in \R^5|\;\; \xi\cdot \xi=0,\;\; \xi^{0}>0 \right\rbrace$, which plays the role of energy-momentum parameter \cite{brmo03,higgs,brmo}.
 
In ambient space of dS universe, the action of \textit{mmc} scalar and spinor free fields is \cite{77,bagamota,higgs}:
 \b \label{conf spinor massless action} S(\Psi,\Phi)=\int \ud\mu(x){\cal L}_{free}(\Psi , \Phi)=\int \ud\mu(x)\left[H \bar{\Psi} \gamma^4\left( -i\barra{x} \barra{\partial}^\top+2i\pm\nu\right) \Psi+\Phi_m \;\partial^\top\cdot\partial^\top\;\Phi_m \right],\e 
 where $\ud\mu(x)$ is the dS-invariant volume element. The $\Psi$ is spinor field and $\bar{\Psi}=\Psi\dg\gamma^0\gamma^4$ is spinor adjoint field in dS ambient space formalism. The $\gamma^\alpha$ are five-matrices with the properties: 
 \b  \gamma^{\alpha}\gamma^{\beta}+\gamma^{\beta}\gamma^{\alpha}
 =2\eta^{\alpha\beta},\;\;\;\;\gamma^{\alpha\dagger}=\gamma^{0}\gamma^{\alpha}\gamma^{0}.\e
 The ambient gamma matrices $\gamma^\alpha$ are different from Minkowski gamma matrices $\gamma^{'\mu}$. The Minkowski gamma matrices are related with ambient gamma matrices as \cite{bagamota}:
 \b \label{gamma relation}\gamma'^{\mu}=\gamma^{\mu}\gamma^4.\e
 In \eqref{conf spinor massless action}, the $\nu$ is related to dS mass parameter $m^2_{f,\nu}=H^2(2+\nu^2\pm i\nu)$, which in flat space limit reduces to Minkowski mass parameter $m$\cite{bagamota}. Also we have $\barra x=\eta_{\alpha\beta}\gamma^{\alpha} x^{\beta}$ and $\cancel{\partial}^\top=\gamma^\alpha\partial_\alpha^\top=\gamma^\alpha\left(\partial_\alpha+H^2x_\alpha x\cdot\partial\right)$.
 In second term of \eqref{conf spinor massless action}, the $\Phi_m$ is {\it mmc}  scalar field \cite{77}.
 The action \eqref{conf spinor massless action} is invariant under the global U(1) symmetry. 
 By replacing transverse derivative $\partial_\alpha^\top$ with this new derivative $D_\alpha\Psi=(\partial_\alpha^\top+{\cal G}B^\top_\alpha\Phi_m)\Psi $, the action \eqref{conf spinor massless action} is invariant under the following transformation \cite{higgs}:
 \b\label{transformation}
 \Psi\longrightarrow\Psi'=e^{-\frac{1}{2}(Hx\cdot B)^{-2}}\Psi
 \;\;\;,\;\;\;
 \Phi_m\longrightarrow\Phi'_m=\Phi_m+(Hx\cdot B)^{-3}\,.
 \e
 Then the interaction Lagrangian is obtained as \cite{higgs}:
 \b \label{higgs spinor lagrangian} {\cal L}_{int}=-i{\cal G}\;H\bar{\Psi} \gamma^4 \barra{x} \barra{B^\top}\Phi_m\Psi. \e
 Here $B_\alpha$ is an arbitrary 5-vector constant that $B^\alpha B_\alpha=0$ and ${\cal G}$ is the coupling constant that determines the interaction intensity. The  scalar field $\Phi_m$ can be written in terms of the massless conformally coupled ({\it mcc}) scalar field as \cite{higgs,77,khrota,gareta}:
 \b\label{magic}
 \Phi_m(x)= \left[HA\cdot\partial^\top + 2 H^3 A\cdot x\right]\Phi_c(x),\e
where $A_\alpha$ is an arbitrary 5-vector constant. For dimensional compatibility, the dimension of $A$ can be fixed as $H^{-2}$. The \textit{mmc} scalar two-point function can be obtained as \cite{higgs}:
 \b \label{mmc 2point function}
 W_m(x,x')=\frac{iH^4}{8\pi^2}
 \frac{({\cal Z}-3)H^4\left[(A\cdot x)^2 +(A\cdot x')^2+A\cdot x\, A\cdot x' \,{\cal Z}\right]+6H^4A\cdot x A\cdot x'-(1-{\cal Z})H^2A\cdot A }{(1-{\cal Z}+i\epsilon)^3},\e
 where $\cal Z$ is ${\cal Z}(x,x')=-H^2\left(x\cdot x'\right)=1+\dfrac{H^2}{2}(x-x')^2$ \cite{brmo,chta}.

\section{the null curvature limit}\label{flat limit}
In the null curvature limit, the {\it mcc} scalar field and spinor field become their counterpart in the Minkowski space \cite{bagamota,brgamo,brmo}:
\b\label{psi flat limit}
\lim_{H\rightarrow 0}\Psi(x)=\psi(X)
\;,\;\;\;
\lim_{H\rightarrow 0}\bar{\Psi}(x)\gamma^4=\bar{\psi}(X), \;\;\; \lim_{H\rightarrow 0} \Phi_c(x)=\phi(X).
\e
By using \eqref{flat cordinates} it is straightforward to show that:
$$
\lim_{H\rightarrow 0}H\barra{x}=-\gamma^4.\
$$
Since the five-vector $B^\alpha$ is constant, $B^4$ can be chosen as zero. Therefore in the null curvature limit one can obtain:
\b \label{B flat limit}
\lim_{H\rightarrow 0}\barra B^{\top}=\gamma^4 B_4+\gamma^\mu B_\mu= B_{\mu}\gamma^{\mu} =-B_{\mu}\gamma^{'\mu}\gamma^4\;;\;\;\;\;\;\mu =0,1,2,3.\e
From \eqref{magic}, one can obtain the $\Phi_m$ in limit of $H\rightarrow 0$ as:
\b \label{phi mmc flat limit}
\lim_{H\rightarrow 0}\Phi_m(x) =0\;.
\e
By using the null curvature limit of:
$$
\lim_{H\rightarrow 0}(x-x')=(X-X'),\;\;\;\lim_{H\rightarrow 0}{\cal Z}=1, \;\; \lim_{H\rightarrow 0}\left(H\;A\cdot x\right)=-A^4 \equiv 0,
$$
one can obtain:
\b \label{mmc 2point flat limit}
\lim_{H\rightarrow 0}{\cal W}_{mmc}=\frac{H^2}{2\pi^2}\;
\frac{A\cdot A }{(X-X)^4}=0.\e
The phase transformations \eqref{transformation} and the covariant derivative $D_\alpha$ in the null curvature limit become:
\b\label{flat limit of transformation}
\psi\longrightarrow\psi'=e^{-\frac{i}{2}(B_4)^{-2}}\psi\;\;,\;\;D_\mu=\partial_\mu.
\e 
Therefore by using relations \eqref{higgs spinor lagrangian} and \eqref{phi mmc flat limit} the null curvature limit of ${\cal L}_{int}$ is:
\b \label{newint}
\lim_{H\rightarrow 0} {\cal L}_{int}=0.
\e

\section{conclusion}\label{Conclusion}
Interaction of spinor field with {\it mmc} scalar field is investigated in ambient space formalism of dS universe. Since the local phase transformation of the dS-spinor field in the null curvature limit become a constant phase transformation the interaction disappeared and then it is seen the interaction Lagrangian and two-point function vanished in this limit. The \textit{mmc} scalar field in dS space-time disappear in the null curvature limit and then it is expected, this field may be a part of a gravitational field. 
\\
{\bf{Acknowledgments}}:  The author wish to express his particular thanks to  Prof. M. V. Takook for useful guidances and consultations and also to M. Dehghani,  M. Rastiveis, R. Raziani and S. Tehrani-Nasab for useful discussions.

\end{document}